\catcode`@=11
\long\def\@makecaption#1#2{
   \vskip 10pt
   \setbox\@tempboxa\hbox{{\small\bf #1.} \ {\small #2}}
   \ifdim \wd\@tempboxa >\hsize       % IF longer than one line:
   {\small\bf #1.} \ {\small #2}\par  % THEN set as ordinary paragraph.
   \else                              %   ELSE  center.
        \hbox to\hsize{\hfil\box\@tempboxa\hfil}
   \fi}
\catcode`@=12

\catcode`@=11
\def\secteqno{\@addtoreset{equation}{section}%
\def\theequation{\thesection.\arabic{equation}}}
\def\endsecteqno{\def\theequation{\@ifundefined{chapter}%
{\arabic{equation}}{\thechapter.\arabic{equation}}}}
\newcounter{subequation}
\def\thesubequation{\alph{subequation}}
\def\sneqnarray{\stepcounter{equation}\let\@currentlabel=\theequation
\setcounter{subequation}{1}
\def\@eqnnum{{\rm (\theequation\thesubequation)}}
\global\@eqcnt\z@\tabskip\@centering\let\\=\@eqncr\let\@@eqncr=\@@sneqncr
$$\halign to \displaywidth\bgroup\@eqnsel\hskip\@centering
 $\displaystyle\tabskip\z@{##}$&\global\@eqcnt\@ne
 \hskip 2\arraycolsep \hfil${##}$\hfil
 &\global\@eqcnt\tw@ \hskip 2\arraycolsep
$\displaystyle\tabskip\z@{##}$\hfil
tabskip\@centering&\llap{##}\tabskip\z@\cr}
\def\endsneqnarray{\@@sneqncr\egroup $$\global\@ignoretrue}
\def\@@sneqncr{\let\@tempa\relax
   \ifcase\@eqcnt \def\@tempa{& & &}\or \def\@tempa{& &}
   \else \def\@tempa{&}\fi
     \@tempa \if@eqnsw\@eqnnum\stepcounter{subequation}\fi
     \global\@eqnswtrue\global\@eqcnt\z@\cr}
\def\nobiblabels{\def\@lbibitem[##1]##2{\@bibitem{##2}}}
\catcode`@=12

     \def\dag{\dagger}
    
\def\bnabla{{\bm \nabla}}

\documentclass[a4paper,11pt]{article}
\pdfoutput=1
\usepackage{jheppub} % for details on the use of the package, please
\usepackage{bm,amssymb,slashed,graphicx,multirow,soul,mathtools,xspace,array}  
\usepackage[german,english]{babel}
\usepackage{hyperref}
\usepackage{slashed}
\usepackage{epsfig}
\usepackage{bbm}
\usepackage{bbold}
\usepackage{xcolor}
\usepackage{tensor}

\begin{document}

\title{Effect of continuum states on the double-heavy hadron spectra}

\arxivnumber{}
\author[1]{Jaume Tarr\'us Castell\`a}
\emailAdd{jtarrus@icc.ub.edu}
\affiliation[1]{Departament de F\'\i sica Qu\`antica i Astrof\'\i sica and Institut de Ci\`encies del Cosmos, 
Universitat de Barcelona, Mart\'\i $\;$ i Franqu\`es 1, 08028 Barcelona, Catalonia, Spain}

\abstract{We present the leading order coupling of double-heavy hadrons to heavy hadron pairs in Born-Oppenheimer effective field theory. We obtain the expressions for the contribution of heavy hadron pairs to the masses and widths of double-heavy hadrons. We apply our result for the specific case of the coupling of the lowest lying heavy hybrids and $D_{(s)}^{(*)}\bar{D}_{(s)}^{(*)}(B_{(s)}^{(*)}\bar{B}_{(s)}^{(*)})$ obtaining a set of selection rules for the decays. We build a model for the coupling potential and compute the corresponding decay widths and the contributions to the mass of the heavy hybrids. We compare our results with the experimental exotic quarkonium spectrum and discuss the most likely experimental candidates for quarkonium hybrids.}

\maketitle
\flushbottom

\section{Introduction}\label{s:int}

In the past two decades we have seen the discovery of numerous heavy exotic hadrons~\cite{Brambilla:2019esw}. These states do not fall in the traditional quark model classification of mesons and baryons as $q\bar{q}$ and $qqq$ states, respectively. A large amount of exotic heavy hadrons have been discovered in the charmonium and bottomonium spectra close and above the heavy meson-antimeson thresholds. The lack of a comprehensive and QCD based description of these states shows a significant gap in our understanding of the strong interactions.

The charmonium and bottomonium spectra can be determined from QCD following a two step process inspired by the Born-Oppenheimer approximation~\cite{Born:1927rpw}. The first one is to determine the static energies in the limit of static heavy quarks. These static energies depend on the dynamics of the light degrees of freedom, gluons and light quarks, and should be obtained from lattice QCD. In a second step, the spectrum is computed using the static energies as potentials for the heavy quark pair. The spectrum of static energies for a heavy quark-antiquark pair obtained with quenched lattice QCD~\cite{Juge:2002br,Bali:2003jq,Capitani:2018rox,Schlosser:2021wnr} shows a ground state corresponding to the conventional quarkonium spectrum and a set of excited states which support a spectrum of hybrid quarkonium states. Furthermore, direct unquenched lattice QCD~\cite{HadronSpectrum:2012gic,Cheung:2016bym,Ryan:2020iog} computations of the quarkonium spectrum also have found the existence of hybrid states roughly consistent with the ones obtained with the Born-Oppenheimer approximation. 

Beyond the quenched approximation, the spectrum of static energies becomes populated by additional states such as heavy meson-antimeson pairs and states containing light-quark hadrons. These open the possibility of transitions and decays for conventional and hybrid quarkonium. Experimental measurements show that some of the states above open flavor thresholds have significant branching ratios for decays to heavy meson-antimeson pairs, however there are some intriguing exceptions. Moreover, some of the exotic states appear to be extremely close to heavy meson thresholds, which has prompted their interpretation as heavy-meson shallow bound states~\cite{Braaten:2003he,Esposito:2016noz,Guo:2017jvc}. Therefore, it is crucial to elucidate the couplings of the Born-Oppenheimer states to the open flavor thresholds and to determine if they are narrow enough to be observed.

An analogous situation is found in other double-heavy hadron sectors. For instance, in the Born-Oppenheimer approach the spectrum of double-heavy baryons has been computed~\cite{Soto:2020pfa,Soto:2021cgk} and a large number of states have been predicted above the $\Lambda$ and heavy meson threshold. However, the stability of such states has not yet been studied. Finally, several tetraquark and pentaquark states have been observed near heavy hadron pair thresholds.

The effective field theory framework that incorporates the heavy-quark mass expansion and the adiabatic expansion between the heavy quarks and the light degrees of freedom is called Born-Oppenheimer effective field theory (BOEFT)~\cite{Brambilla:2017uyf,Soto:2020xpm}. This effective field theory (EFT) has been used to study numerous properties of double-heavy hadrons: the hybrid quarkonium spectrum~\cite{Berwein:2015vca,Oncala:2017hop} including hyperfine splittings~\cite{Brambilla:2018pyn,Brambilla:2019jfi,Soto:2023lbh}, hybrid quarkonium semi-inclusive transitions~\cite{Oncala:2017hop,TarrusCastella:2021pld,Brambilla:2022hhi}, conventional and hybrid quarkonium (di)pion transitions~\cite{Pineda:2019mhw,TarrusCastella:2021pld} and the double-heavy baryon spectrum~\cite{Soto:2020pfa} including hyperfine splittings~\cite{Soto:2021cgk}.

The objective of the present article is to present the couplings of double-heavy hadrons to pairs of heavy hadrons in BOEFT and to compute the corresponding decay widths and mass contributions. As an example we will discuss the coupling of the lowest lying heavy hybrids and $D_{(s)}^{(*)}\bar{D}_{(s)}^{(*)}(B_{(s)}^{(*)}\bar{B}_{(s)}^{(*)})$. In order to obtain numerical estimates for this case, we will propose a model for the coupling potential. 

A closely related problem is that of conventional quarkonium mixing with and decaying to heavy meson-antimeson pairs. This problem has been studied in Refs.~\cite{Bicudo:2019ymo,Bicudo:2020qhp,Bicudo:2022ihz,Bruschini:2020voj,Bruschini:2021sjh,Bruschini:2023zkb} using models based on an extension of the Born-Oppenheimer approximation, as well as using BOEFT in Ref.~\cite{TarrusCastella:2022rxb}. In the present article we generalize the formalism presented in Ref.~\cite{TarrusCastella:2022rxb} to the coupling of double-heavy hadrons to pairs of heavy hadrons in Section~\ref{s:ths}. As an example of our formalism, in Section~\ref{s:hqdm}, we study the specific case of quarkonium hybrids coupled to heavy meson-antimeson pairs.

\section{Continuum states in BOEFT}\label{s:ths}

Let us denote the double-heavy hadron fields by $\Psi_{\kappa\alpha}(t,\,\bm{r},\,\bm{R})$, with $\bm{r}$ and $\bm{R}$ being the relative and center-of-mass coordinates of the heavy quark pair, respectively. The subindex $\kappa$ indicates the spin of the light degrees of freedom, gluons and light quarks, and $\alpha=-\kappa,..,\kappa$ is its component in the standard spherical basis. The light degrees of freedom are also characterized by parity, $p$, and possibly, charge conjugation, $c$, which we will not track explicitly in our notation unless we refer to an specific example. The fields $\Psi_{\kappa}$ also contain the heavy-quark spin state, however as we will work at leading order in the heavy-quark mass expansion, tracking the corresponding indices will not be necessary.

Similarly, we represent the heavy hadron pairs by ${\cal M}_{\kappa\alpha}(t,\,\bm{r},\,\bm{R})$. In this case $\kappa$ is any of the possible combinations of the light degrees of freedom spin of each one of the hadrons forming the pair. Note that, when the heavy quark spin is taken into account, a ${\cal M}_{\kappa}$ field can be decomposed into a linear combination of physical heavy hadron pair states and that each one of these pairs can appear in more than one field. Both the $\Psi$ and ${\cal M}$ fields carry light-quark flavor quantum numbers, which for the purpose of this work must be equal. We do not show them explicitly since they do not affect the construction of the EFT. However, it should be kept in mind that the matching coefficients, such as the potentials, do depend on them.

The leading order Lagrangian in the $1/m_Q$ expansion, with $m_Q$ the heavy-quark mass, except for the relative coordinate kinetic term, for one double-heavy hadron field and one heavy hadron pair field is as follows:
\begin{align}
{\cal L}=&\int d^3\bm{R}\,d^3\bm{r}\left\{\Psi^\dag_\kappa\left[i\partial_t+\frac{\bnabla_r^2}{m_Q}-V_{\Psi_\kappa}^{(0)}(\bm{r})\right]\Psi_\kappa+{\cal M}^\dag_{\kappa'}\left[i\partial_t+\frac{\bnabla_r^2}{m_Q}-V_{{\cal M}_{\kappa'}}^{(0)}(\bm{r})\right]{\cal M}_{\kappa'}\right\}\,.\label{boeft}
\end{align}

The symmetry group of two static heavy quarks is $D_{\infty h}$, which is a cylindrical symmetry group. The representations of $D_{\infty h}$ are labeled as $\Lambda^\sigma_{\eta}$. $\Lambda$ is the absolute value of the projection of the light degrees of freedom spin on the axis joining the two heavy quarks, given by $\hat{\bm{r}}$. It is traditionally represented by capital
Greek letters, $\Sigma,\Pi,\Delta,\Phi...$ for $\Lambda=0,1,2,3...$ The quantum number $\eta=\pm 1$ is the $P$ or $CP$ eigenvalue, for heavy quark-quark and heavy quark-antiquark, respectively. It is denoted by $g = + 1$ and $u = - 1$. Additionally for $\Sigma$ representations there is a symmetry under reflection with respect to a plane passing through the axis $\hat{\bm{r}}$. The eigenvalues of the corresponding symmetry operator are $\sigma=\pm 1$ and are indicated as superscript. The static potential for the double-heavy hadrons can be expanded in representations of this symmetry group~\cite{Soto:2020xpm}
\begin{align}
V_{\Psi_\kappa}^{(0)}(\bm{r})=&\,\sum_{\Lambda=0}^{|\kappa|}V_{\kappa\Lambda}^{(0)}(r)\sum_{\lambda=\pm\Lambda}P_{\kappa\lambda}(\hat{\bm{r}})P^\dag_{\kappa\lambda}(\hat{\bm{r}})\,,\label{edihsp}
\end{align}
where we have introduced the projection vectors $P_{\kappa\lambda}(\hat{\bm{r}})$. Notice that, the potential in Eq.~\eqref{edihsp} is a matrix in the light degrees of freedom spin space. The projection vectors project the fields from a spherical spin representation to $D_{\infty h}$ ones along the heavy quark pair axis $\hat{\bm{r}}$. Following Ref.~\cite{Brambilla:2017uyf}, they are defined as
\begin{align}
\Psi_{\kappa\alpha}&=\sum_{\lambda=-\kappa,..,\kappa}\left(P_{\kappa\lambda}(\hat{\bm{r}})\right)_\alpha \Psi_{\kappa\lambda}\,,\label{pb1}\\
\Psi_{\kappa\lambda}&=\left(P^*_{\kappa\lambda}(\hat{\bm{r}})\right)^\alpha \Psi_{\kappa\alpha}\,,\label{pb2}
\end{align}
and, furthermore, they must fulfill
\begin{align}
&\bm{S}^2_\kappa P_{\kappa\lambda}(\hat{\bm{r}})=\kappa(\kappa+1) P_{\kappa\lambda}(\hat{\bm{r}})\,,\\
&\left(\hat{\bm{r}}\cdot \bm{S}_{\kappa}\right) P_{\kappa\lambda}(\hat{\bm{r}})=\lambda P_{\kappa\lambda}(\hat{\bm{r}})\,.
\end{align}
The projection vectors can be identified as the spin functions in the helicity basis~\cite{Khersonskii:1988krb}, therefore, a general expression for their components can be given in terms of Wigner D-functions
\begin{align}
\left(P_{\kappa\lambda}(\hat{\bm{r}})\right)_\alpha=D^\kappa_{-\lambda-\alpha}(0,\hat{\bm{r}})\,.
\end{align}

In the $D_{\infty h}$ basis in Eqs.~\eqref{pb1}-\eqref{pb2} the static potential $V_{\Psi_\kappa}^{(0)}$ is diagonal~\cite{Brambilla:2017uyf}, however, due to the dependence of the projection vectors on the angles of $\hat{\bm{r}}$, in this basis the kinetic operator becomes non-diagonal. These off-diagonal terms are the so-called nonadiabatic coupling which induces a mixing between the $\Psi_{\kappa\lambda}$ components. A more in-depth discussion can be found in Refs.~\cite{Berwein:2015vca,Brambilla:2017uyf,Oncala:2017hop,Soto:2020pfa}. In Refs.~\cite{Bruschini:2020voj,Bruschini:2021sjh,Bruschini:2023zkb,Bruschini:2023tmm}, the basis in terms of spherical spin components is referred as diabatic basis, while the $D_{\infty h}$ basis is called adiabatic basis.

The static potential between the heavy hadron pairs is not known in general. The static energies between heavy meson-meson and heavy meson-antimeson pairs has been studied on the lattice in Refs.~\cite{Bali:2005fu,Bicudo:2012qt,Brown:2012tm,Bicudo:2015vta,Bicudo:2015kna,Prelovsek:2019ywc,Bulava:2019iut,Bicudo:2021qxj,Sadl:2021bme}. The results show that the static energies are mostly flat lines at the energy corresponding to the heavy meson masses except in some cases in the short-distance limit where the potential is attractive or repulsive depending on the specific heavy-meson pair. However, only in Refs.~\cite{Bali:2005fu,Bulava:2019iut} the coupled system of the heavy meson-antimeson pair and quarkonium static energies has been studied. The analysis of Ref.~\cite{TarrusCastella:2022rxb} indicates that when the coupling is taken into account the static potential of the heavy meson-antimeson pair becomes flat. Hence, we are going to assume that  
\begin{align}
V_{{\cal M}_{\kappa'}}^{(0)}(\bm{r})=\left(m^{(0)}_T-2m_Q\right)\mathbb{1}_{\kappa'}\,,
\end{align}
with $m^{(0)}_T$ being the total mass of the heavy hadron pair at leading order in the heavy-quark mass expansion, and $\mathbb{1}_{\kappa'}$ is an identity in the spin-$\kappa'$ space. Naturally, at long distances one should expect some deviation from this picture due to light-quark meson exchanges, but it will not be taken into account. Likewise, we do not consider the width of the heavy hadrons. As the heavy hadron pairs are non-interacting they form a continuum of states.

The BOEFT operator coupling a $\Psi_{\kappa}$ field with a heavy hadron pair field ${\cal M}_{\kappa'}$ is as follows:
\begin{align}
\mathcal{L}_{\kappa'\kappa}=-\int d^3\bm{R}\,d^3\bm{r}\sum_{\Lambda}V_{\kappa'\kappa\Lambda}(r)\sum_{\lambda=\pm\Lambda}\left\{{\cal M}^{\dag\alpha}_{\kappa'} (P_{\kappa'\lambda}(\bm{\hat r}))_\alpha(P^*_{\kappa \lambda}(\bm{\hat r}))^{\beta}\Psi_{\kappa\beta}+{\rm h.c.}\right\}\,,\label{mix_lgrngn}
\end{align}
with the sum over $\Lambda$ going over the range $(0,...,{\rm min}(|\kappa|,|\kappa'|))$\footnote{For fractional spin the sum starts at $1/2$.}. It is interesting to write Eq.~\eqref{mix_lgrngn} in terms of the projected field basis in Eqs.~\eqref{pb1}-\eqref{pb2}. In this basis, the coupling becomes trivial:
\begin{align}
\mathcal{L}_{\kappa'\kappa}=-\int d^3\bm{R}\,d^3\bm{r}\sum_{\Lambda}V_{\kappa'\kappa\Lambda}(r)\sum_{\lambda=\pm\Lambda}\left\{{\cal M}^{\dag}_{\kappa'\lambda} \Psi_{\kappa\lambda}+{\rm h.c.}\right\}\,.\label{mix_lgrngn2}
\end{align}
In this form it is clear that two BOEFT fields are coupled at leading order if they belong to the same $D_{\infty h}$ representation. 

The double-heavy hadron states are as follows:
\begin{align}
|n,\,\ell,\,m_\ell\,,\kappa\rangle&=\int d^3\bm{R}\,d^3\bm{r}\,\Psi^{(n\ell)}_{\kappa\alpha}(\bm{r}) \,\Psi^{\dag \alpha}_{\kappa}(\bm{r},\,\bm{R})|0\rangle\,,\label{state_sph}
\end{align}
with 
\begin{align}
\Psi^{(n\ell)}_{\kappa\alpha}(\bm{r})&=\sum^{\ell+\kappa}_{l=|\ell-\kappa|}\psi_{n\ell l}(r)\left(Y^{l\kappa}_{\ell m_\ell }(\hat{\bm{r}})\right)_\alpha\,,\label{wf_dhh}
\end{align}
the wave function solution of the following coupled Schrödinger equations
\begin{align}
\left[-\frac{\bnabla^2_r}{m_Q}\delta_{\alpha\beta}+{\left(V_{\Psi_\kappa}^{(0)}(\bm{r})\right)_{\alpha\beta}}\right]\Psi^{(n\ell)}_{\kappa\beta}(\bm{r})&=E^{(0)}_{n\ell}\Psi^{(n\ell)}_{\kappa\alpha}(\bm{r})\,,
\end{align}
and $n$ is the principal quantum number. The components of the tensor spherical harmonics~\cite{Khersonskii:1988krb} defined as 
\begin{align}
\left(Y^{l\kappa}_{\ell m_\ell }(\hat{\bm{r}})\right)_\alpha={\cal C}^{\ell m_\ell}_{l m_l\,\kappa -\alpha}(-1)^{\kappa-\alpha} Y_{lm_l}(\hat{\bm{r}})\,,\label{dhh_state}
\end{align}
with ${\cal C}$ denoting a Clebsch-Gordan coefficient. The tensor spherical harmonics combine the spin of the light degrees of freedom ($\kappa$) and the heavy-quark pair angular momentum ($l$) and depend on the total angular momentum of the double-heavy hadron ($\ell$). The sum over $l$ in Eq.~\eqref{wf_dhh} only runs over even or odd values depending on the parity of the state.

The continuum states can be defined analogously. The radial wave function corresponds to the projection of a plane wave into an angular momentum $l$ and is independent of $\ell$. Nevertheless, it will be convenient to use continuum states with definite $l$ and $\ell$
\begin{align}
|k,\,\ell,\,m_\ell\,,l\,,\kappa\rangle&=\int d^3\bm{R}\,d^3\bm{r}\,\left(4\pi i^{-l}j_l(kr)\right)\left(Y^{l\kappa}_{\ell m_\ell }(\hat{\bm{r}})\right)_\alpha\,{\cal M}^{\dag \alpha}_{\kappa}(\bm{r},\,\bm{R})|0\rangle\,,\label{state_cont}
\end{align}
with, $k=|\bm{k}|$, the absolute value of the relative momentum between the heavy hadron pair and $j_l(kr)$ a spherical Bessel function.

Now, let us work out the states in the $D_{\infty h}$ basis defined by the transformations in Eq.~\eqref{pb1} and \eqref{pb2}. This can be done  by introducing the identity $\mathbb{1}_\kappa=\sum_{\lambda}P_{\kappa\lambda}P^\dag_{\kappa\lambda}$ and using
\begin{align}
\left(Y^{l\kappa}_{\ell m_\ell }(\hat{\bm{r}})\right)_\alpha(P^*_{\kappa \lambda})^{\alpha}&=(-i)^\kappa\sqrt{\frac{2\ell+1}{4\pi}}{\cal C}^{l 0}_{\ell \lambda\,\kappa -\lambda}D^\ell_{\lambda-m_\ell}(0,\hat{\bm{r}})\,,
\end{align}
which follows from Wigner D-function identities. We arrive to
\begin{align}
|n,\,\ell,\,m_\ell,\,\kappa\rangle&=(-i)^\kappa\sum_\lambda\int d^3\bm{R}\,d^3\bm{r}\,\phi^{(\kappa\lambda)}_{n\ell}(r)\sqrt{\frac{2\ell+1}{4\pi}}D^\ell_{\lambda-m_\ell}(0,\hat{\bm{r}})\Psi^{\dag}_{\kappa\lambda}(\bm{r},\,\bm{R})|0\rangle\,,\label{cyl_state}
\end{align}
with
\begin{align}
&\phi^{(\kappa\lambda)}_{n\ell}(r)=\sum^{\ell+\kappa}_{l=|\ell-\kappa|}{\cal C}^{l 0}_{\ell \lambda\,\kappa -\lambda}\psi_{n\ell l}(r)\,.\label{basis_rel}
\end{align}
An analogous expression to Eq.~\eqref{cyl_state} can be derived for the continuum states $|k,\,\ell,\,m_\ell\,,l\,,\kappa\rangle$. Eq.~\eqref{basis_rel} relates the radial wave functions for hybrid quarkonium from Ref.~\cite{Oncala:2017hop} to the ones in Ref.~\cite{Berwein:2015vca}\footnote{The orbital wave functions of Ref.~\cite{Berwein:2015vca} are $v^\lambda_{\ell m_\ell}(\hat{\bm{r}})=\sqrt{\frac{2\ell+1}{4\pi}}D^\ell_{-\lambda-m_\ell}(0,\hat{\bm{r}})$.}.

Now, we are in a position to easily compute the expected value of the coupling operator in the Lagrangian in Eq.~\eqref{mix_lgrngn2} for the transition of a double-heavy hadron state into a pair of heavy hadrons with angular momentum $l'$ and relative momentum $k$. Using the orthogonality of the Wigner D-functions we find
\begin{align}
\langle k,\,\ell',\,m_{\ell'},\,l',\,\kappa'|\sum_{\lambda=\pm\Lambda}\int d^3\bm{r}{\cal M}^{\dag\alpha}_{\kappa'\lambda}V_{\kappa'\kappa\Lambda}(r)\Psi_{\kappa\lambda}|n,\,\ell,\,m_\ell,\,\kappa\rangle &\nonumber\\
=\delta_{\ell'\ell }\delta_{m_{\ell^\prime}m_{\ell}} 4\pi (-i)^{(\kappa-\kappa'-l')}a^{\kappa'\kappa\Lambda}_{n \ell; l'}(k)&\,,\label{phase_out}
\end{align}
with
\begin{align}
a^{\kappa'\kappa\Lambda}_{n\ell;l'}(k)=\sum_{\lambda=\pm\Lambda}{\cal C}^{l' 0}_{\ell \lambda\,\kappa' -\lambda}\int dr r^2 j_{l'}(kr)V_{\kappa'\kappa\Lambda}(r)\phi^{(\kappa\lambda)}_{n\ell}(r)\label{v_factor}
\end{align}
The allowed values of the angular momentum of the heavy hadron pair are the even or odd values in the range $(|\ell-\kappa'|,...,\ell+\kappa')$ depending on the parity of the double-heavy hadron state.

Let us compute the continuum states contribution to the double-heavy hadron self-energy
\begin{align}
&\Sigma^{\kappa'\kappa\Lambda}_{n\ell;l'}(E)=\frac{4\mu}{\pi}\int dk k^2\frac{[a^{\kappa'\kappa\Lambda}_{n\ell; l'}(k)]^2}{p(E)^2-k^2}\,,\label{self_en}
\end{align}
where $p(E)=\sqrt{2\mu (E+i\epsilon+2m_Q-m_T)}$ with $\mu$ and $m_T$ being the heavy hadron pair reduced and total masses, respectively. In order to take into account the self-energy to all orders in perturbation theory, we can resum the self-energy into the double-heavy hadron propagator. Using Sokhotsky’s formula we separate the real and imaginary parts of the self-energy and find the bound state energy of the double-heavy hadron as the roots of
\begin{align}
E^{(r)}_{n\ell}-E_{n\ell}-\sum_{\kappa'\Lambda\,l'}{\cal P}(\Sigma^{\kappa'\kappa\Lambda}_{n\ell;l'}(E^{(r)}_{n\ell}))=0\,,
\end{align}
where ${\cal P}$ denotes Cauchy's principal part and the sum goes over all the possible intermediate heavy hadron pair states. The energy shift and width can be obtained as
\begin{align}
&\delta E^{\kappa'\kappa\Lambda}_{n\ell;l'} = {\cal P}(\Sigma^{\kappa'\kappa\Lambda}_{n\ell;l'}(E^{(r)}_{n\ell}))\,,\\
&\Gamma^{\kappa'\kappa\Lambda}_{n\ell;l'}=4\mu\, p_r[a^{\kappa'\kappa\Lambda}_{n\ell; l'}(p_r)]^2\,,
\end{align}
with $p_r=p(E^{(r)}_{n\ell})$.

For the rest of the paper we will use $E_{n\ell}=E^{(0)}_{n\ell}$. However, one of the strengths of our approach is that we can improve the accuracy of this quantity by including higher order terms in BOEFT. This is important, because when $E_{n\ell}+2m_Q-m_T$ is close to zero, i.e when the double-heavy hadron state is close to a heavy hadron pair threshold, the contribution of the continuum states is very sensitive to small variations of $E_{n\ell}$.

The conventional quarkonium case, which was studied in Ref.~\cite{TarrusCastella:2022rxb}, corresponds to setting $\kappa^{pc}=0^{++}$ in the expressions of this section. In this case $\ell$ must be equal to $l$. Furthermore, $\lambda=0$ as it is the projection of $\kappa$, which corresponds to the $\Sigma_g^+$ representation of the static energy of conventional quarkonium. The wave function $\phi^{(\kappa\lambda)}_{n\ell}(r)$ in Eq.~\eqref{basis_rel} reduces to the radial wave function of conventional quarkonium. To match Eq.~\eqref{phase_out}-\eqref{v_factor} with $\kappa^{pc}=0^{++}$  to the results of Ref.~\cite{TarrusCastella:2022rxb} a $(-1)^{\kappa'}$ phase has to be shifted from the definition of Eq.~\eqref{phase_out} to Eq.~\eqref{v_factor}.

\section{Hybrid quarkonium and heavy-meson pairs}\label{s:hqdm}

Let us now apply the results of the previous section to study the effect of the heavy meson-antimeson thresholds to the hybrid quarkonium spectrum. In table~\ref{qhreps} we show the quantum numbers of the four lowest mass gluelumps~\cite{Herr:2023xwg} and the $D_{\infty h}$ representations of the corresponding static states~\cite{Brambilla:1999xf,Berwein:2015vca}. Each heavy meson-antimeson pair is characterized by the spin and parity of the light-quark states forming the two heavy mesons, which we label as $\kappa_1^{p_1}$ and $\kappa_2^{p_2}$ in table~\ref{hmpreps}. Combining the spin and parity of the light quarks we find their total $\kappa^{pc}$. From these, we work out the possible $\Lambda^\sigma_\eta$ representations in table~\ref{hmpreps}. A hybrid (or conventional) quarkonium state couples to a heavy meson-antimeson pair at leading order in the heavy-quark mass expansion if they can be projected into the same $D_{\infty h}$ representation. For the lowest lying hybrid quarkonium, with $\kappa=1^{+-}$ gluelump, the only matching $D_{\infty h}$ representation with the lowest lying heavy meson-antimeson pairs, i.e. $D_{(s)}^{(*)}\bar{D}_{(s)}^{(*)}(B_{(s)}^{(*)}\bar{B}_{(s)}^{(*)})$, is $\Sigma_u^-$ corresponding to a $0^{-+}$ light-quark state. For this case the form factor in Eq.~\eqref{v_factor} takes the following simple form
\begin{align}
a^{0^{-+}1^{+-}\Sigma}_{n\ell;l'}(k)&=\delta_{\ell l'}\int dr r^2 j_{l'}(kr) V_{01\Sigma}(r) \phi^{(10)}_{n\ell}(r)\,.\label{hh_hmp}
\end{align}

\begin{table}[ht]
\centering
\begin{tabular}{cc} \hline\hline
$\kappa^{pc}$ & $D_{\infty h}$                  \\ \hline
$0^{++}$      & $\Sigma_g^+$                    \\ \hline
$1^{+-}$      & $\Sigma_u^-,\,\Pi_u$            \\ 
$1^{--}$      & $\Sigma_g^+,\,\Pi_g$            \\                            
$2^{+-}$      & $\Sigma_u^-,\,\Pi_u,\,\Delta_u$ \\ 
$2^{--}$      & $\Sigma_g^+,\,\Pi_g,\,\Delta_g$ \\ \hline\hline
\end{tabular}
\caption{Total spin ($\kappa$), parity ($p$) and charge conjugation ($c$) of the glue state in hybrid quarkonium and the corresponding allowed $D_{\infty h}$ representations resulting from projecting the spin into the heavy quark-antiquark axis. The first entry corresponds to conventional quarkonium, which containing no gluonic degrees of freedom, corresponds to $0^{++}$.}
\label{qhreps}
\end{table}

\begin{table}[ht]
\centering
\begin{tabular}{c|c|c|c}  \hline\hline
$\kappa_1^{p_1}$ & $\kappa_2^{p_2}$ & $\kappa^{pc}$ & $D_{\infty h}$ \\ \hline
$(1/2)^+$    & $(1/2)^+$        & $0^{-+}$ & $\Sigma_u^-$ \\
             &                  & $1^{--}$ & $\Sigma_g^+,\,\Pi_g$ \\ \hline
$(1/2)^+$    & $(1/2)^-$        & $0^{++}$ & $\Sigma_g^+$\\                                 
             &                  & $1^{+-}$ & $\Sigma_u^-,\,\Pi_u$ \\
$(1/2)^+$    & $(3/2)^-$        & $1^{+-}$ & $\Sigma_u^-,\,\Pi_u$  \\
             &                  & $2^{++}$ & $\Sigma_g^+,\,\Pi_g,\,\Delta_g$\\ \hline
$(1/2)^-$    & $(3/2)^-$        & $1^{--}$ & $\Sigma_g^+,\,\Pi_g$  \\
             &                  & $2^{-+}$ & $\Sigma_u^-,\,\Pi_u,\,\Delta_u$ \\ 
$(1/2)^-$    & $(1/2)^-$        & $0^{-+}$ & $\Sigma_u^-$ \\
             &                  & $1^{--}$ & $\Sigma_g^+,\,\Pi_g$ \\
$(3/2)^-$    & $(3/2)^-$        & $0^{-+}$ & $\Sigma_u^-$ \\
             &                  & $1^{--}$ & $\Sigma_g^+,\,\Pi_g$ \\
             &                  & $2^{-+}$ & $\Sigma_u^-,\,\Pi_u,\,\Delta_u$ \\
             &                  & $3^{--}$ & $\Sigma_g^+,\,\Pi_g,\,\Delta_g,\,\Phi_g$\\ \hline\hline
\end{tabular}
\caption{Total spin ($\kappa$), parity ($p$) and charge conjugation ($c$) of the light-quark state of heavy meson-antimeson pairs and their projections into $D_{\infty h}$ representations. The $\kappa^{pc}$ are obtained as all the possible combinations of the spin of each heavy meson light-quark state. Note that, the representation corresponding to the light-quark state of the heavy antimeson in the pair should be taken conjugated. The ground state heavy meson doublet corresponds to $(1/2)^+$ while  $(1/2)^-$ and $(3/2)^-$ correspond to excited heavy meson states. }
\label{hmpreps}
\end{table}

We can infer the following selection rules for the decays of hybrid quarkonium into $D_{(s)}^{(*)}\bar{D}_{(s)}^{(*)}(B_{(s)}^{(*)}\bar{B}_{(s)}^{(*)})$ from Eq.~\eqref{hh_hmp}. First, we can see that only states with a $\lambda=0$ component can decay through this channel, as has been recently pointed out in Ref.~\cite{Bruschini:2023tmm}. In table~\ref{multi} we summarize the details of the $\kappa=1^{+-}$ gluelump hybrid quarkonium states~\cite{Berwein:2015vca,Oncala:2017hop}. Since the $\lambda=0$ component corresponds to the $\Sigma_u^-$ static state, we can read off table~\ref{multi} which hybrid states are allowed to decay through this channel. Using the notation $l_\ell$, these are $s_0$, $(s\backslash d)_1$ and $(p\backslash f)_2$. On the other hand the states $p_1$ and $d_2$ are forbidden to decay. Higher angular momentum, not shown in table~\ref{multi}, with mixed $l$ are also allowed to decay. It is interesting to note that the mixed states are dominated by the $\Pi_u$ component~\cite{Berwein:2015vca}, which adds a small suppression to the decays into heavy mesons. The second selection rule is that the partial wave of the heavy-meson pair must be equal to the total angular momentum of the hybrid quarkonium states $\ell$. This is a consequence of the light-quark state of the heavy-meson pairs coupling to hybrids being spin $0$. Therefore, only one partial wave is allowed in the decays of quarkonium hybrids into $D^{(*)}\bar{D}^{(*)}(B^{(*)}\bar{B}^{(*)})$ at leading order in BOEFT.

In previous works~\cite{Tanimoto:1982eh,LeYaouanc:1984gh,Iddir:1988jc,Ishida:1991mx,Close:1994hc,Page:1996rj,Page:1998gz,Kou:2005gt}, it had been argued that quarkonium hybrids cannot decay into $S$-wave heavy-meson pairs, while the more recent Ref.~\cite{Bruschini:2023tmm} argued that these decays are not suppressed. Our results show that, in BOEFT, decays into $S$-wave heavy-meson pairs are $1/m_Q$-suppressed except for $\ell=0$ quarkonium hybrids.

\begin{table}[ht]
\centering
\begin{tabular}{ccccc}\hline\hline 
 $\ell$ & $l$             & $\Lambda^\sigma_\eta$        & $j^{pc}(s_{Q\bar{Q}}=\{0\,,1\})$ & Refs.~\cite{Braaten:2014qka,Berwein:2015vca} \\ \hline
 $0$    & $s$             & $\Sigma_u^-$                 & $\{0^{++},1^{+-}\}$              & $H_3$                                       \\
 $1$    & $s\backslash d$ & $\Sigma_u^-\backslash \Pi_u$ & $\{1^{--},(0,1,2)^{-+}\}$        & $H_1$                                       \\
 $1$    & $p$             & $\Pi_u$                      & $\{1^{++},(0,1,2)^{+-}\}$        & $H_2$                                       \\
 $2$    & $p\backslash f$ & $\Sigma_u^-\backslash \Pi_u$ & $\{2^{++},(1,2,3)^{+-}\}$        & $H_4$                                       \\ 
 $2$    & $d$             & $\Pi_u$                      & $\{2^{--},(1,2,3)^{-+}\}$        & $H_5$                                       \\ \hline\hline 
\end{tabular}
\caption{Quantum numbers of hybrid quarkonium with $1^{+-}$ gluelump. The quantum numbers are as follows: $l(l+1)$ is the eigenvalue of $\bm{L}^2_{Q\bar{Q}}$, the square of the angular momentum of the heavy-quark pair. $\ell(\ell+1)$ is the eigenvalue of $\bm{L}^2=(\bm{L}_{Q\bar{Q}}+\bm{S}_1)^2$, with $\bm{S}_1$ the gluelump spin operator. $s_{Q\bar{Q}}(s_{Q\bar{Q}}+1)$ is the eigenvalue of $\bm{S}^2_{Q\bar{Q}}$, the square of the total heavy-quark spin operator. The square of the total spin $\bm{J}^2=(\bm{L}+\bm{S}_{Q\bar{Q}})^2$ has eigenvalue $j(j+1)$. As we are working at leading order in the heavy-quark mass expansion the hybrid quarkonium masses only depend on $\ell$ and a principal quantum number $n$. The set of total spin states for each $\ell$ form degenerate multiplets at this order. The $\Lambda^\sigma_{\eta}$ column indicates the $D_{\infty h}$ representations of static energies that contribute to the states. For $\ell>0$, two sets of solutions with opposite parity exists. One of these corresponds to the mixing the $\Sigma_u^-$ and $\Pi_u$ static states while the other is purely $\Pi_u$\cite{Berwein:2015vca}. The solutions that mix static states also mix the heavy-quark pair angular momentum~\cite{Oncala:2017hop}. The last column indicates the labels for the spin-symmetry multiplets used in Refs.~\cite{Braaten:2014qka,Berwein:2015vca}.}
\label{multi}
\end{table}

The coupling potential between the $\kappa=1^{+-}$ hybrid field and the $\kappa'=0^{-+}$ heavy-meson pair field is a nonperturbative quantity and so far there is no lattice determination of it. As a proof of concept, and in order to obtain numerical estimates for the decay widths and mass contributions, we will build a model for the coupling potential. The model is constructed following Refs.~\cite{Oncala:2017hop,Soto:2021cgk,TarrusCastella:2022rxb,Soto:2023lbh}, that is by interpolating between short- ($r\lesssim 1/\Lambda_{\rm QCD}$) and long-distance ($r\gtrsim 1/\Lambda_{\rm QCD}$) parametrizations. This kind of parametrization was used in Ref.~\cite{TarrusCastella:2022rxb} to fit the lattice data~\cite{Bali:2005fu} for the quarkonium and heavy meson-antimeson pair mixing potential and was found to provide a good description. Therefore, we use this parametrization as inspiration for our model. 

\begin{figure}[ht]
\centering
\begin{tabular}{cc}
\includegraphics[width=0.3\linewidth]{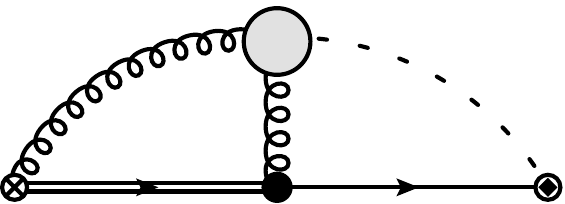} & \includegraphics[width=0.3\linewidth]{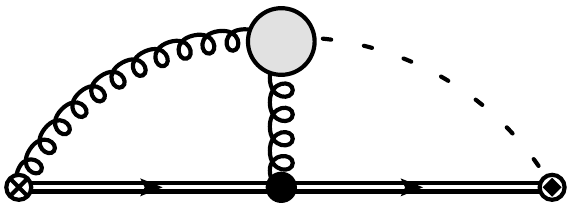} \\
(a) & (b) \\
\end{tabular}
\caption{Leading order diagrams in weakly-coupled pNQRCD contributing to the hybrid quarkonium and heavy-meson pair coupling potential. The diagrams are in coordinate space. The single, double and curly lines represent the heavy-quark singlet, heavy-quark octet and gluon fields respectively. The dashed line corresponds to the light-quark fields. The black dots stand for the dipolar chromoelectric operators. The circles with a cross and with a diamond represent the quarkonium hybrid and heavy meson-antimeson pair states, respectively. It should be noted that while at leading order the quarkonium hybrid only overlaps with the octet field, the heavy-meson pair can overlap both with the singlet and octet fields. The gray circles represent nonperturbative dynamics of the light degrees of freedom.
}
\label{mix_sd}
\end{figure}

The short-distance region is, by construction, the region where the multipole expansion is valid. Therefore, in this region the potential can be expanded in a series in powers of $r$. Such series can be obtained directly using weakly-coupled pNRCQD~\cite{Pineda:1997bj,Brambilla:1999xf}, which is an EFT incorporating the heavy-quark mass and multipole expansions. Several similar examples can be found in Refs.~\cite{Oncala:2017hop,Brambilla:2019jfi,Soto:2021cgk}. This computation yields the coefficients of the expanded potential as correlators of gluon and/or light-quark operators, which encode the nonperturbative dynamics. The two weakly-coupled pNRCQD leading order diagrams contributing to coupling potential are depicted in Fig.~\ref{mix_sd}. We must consider two diagrams because the heavy meason-antimeson pair state can overlap both with the heavy-quark singlet and octet fields. However, in both instances, a single dipolar chromoelectric operator is enough to produce the transition between the quarkonium hybrid and heavy-meson pair. The explicit form of this operators can be found, for instance, in Eq.~(3) of Ref.~\cite{Brambilla:1999xf}. The singlet-to-octet coupling reads as ${\rm Tr}[{\rm S}^\dagger\bm{r}\cdot \bm{{\rm E}} {\rm O}+h.c]$ and the octet-to-octet as ${\rm Tr}[{\rm O}^\dagger\bm{r}\cdot \bm{{\rm E}} {\rm O}+h.c]$. Therefore, at leading order, the short-distance coupling potential must be proportional to $r$: 
\begin{align}
V^{\,{\rm (s.d.)}}_{01\Sigma}(r)=c_s r \,.  
\end{align}

At long distances the coupling potential can be expanded in powers of $1/(\Lambda_{\rm QCD} r)$. Assuming the potential vanishes at infinity only positive powers should be considered. Moreover, the expansion will be dominated by the lowest power, hence only this one will be considered.
\begin{align}
V^{\,{\rm (l.d.)}}_{01\Sigma}(r)=\frac{c_l}{r^{n_l}}\,.  
\end{align}
The long distance quarkonium-heavy-meson pair mixing potential obtained from lattice data was found to behave as $n_l=3$~\cite{TarrusCastella:2022rxb}.

The sizes of $c_s$ and $c_l$ can be estimated from dimensional analysis and the fact that they can only depend on the $\Lambda_{\rm QCD}$ scale.
\begin{align}
c_s\sim \Lambda^2_{\rm QCD}\,,\quad c_l\sim \Lambda^{1-n_l}_{\rm QCD}\,.\label{dim_an}
\end{align}
When a numerical value is needed we will use $\Lambda_{\rm QCD}=0.2$~GeV.

We interpolate between short and long-distance as follows
\begin{align}
V_{01\Sigma}(r)=w_s(r) V^{\,{\rm (s.d.)}}_{01\Sigma}(r)+w_l(r)V^{\,{\rm (l.d.)}}_{01\Sigma}(r)\,,\label{mixpl}
\end{align}
with the interpolating functions $w_s=(r_0/(r+r_0))^{n_l+2}$ and $w_l=(r/(r+r_0))^{n_l+2}$. The $r_0$ parameter determines the value of $r$ where both interpolating functions are equal. We will use values of $r_0\sim 0.25$~fm since it was found in Ref.~\cite{TarrusCastella:2022rxb} to provide a good description of the quarkonium-heavy-meson pair mixing potential obtained from the lattice data of Ref.~\cite{Bali:2005fu} as the lattice mixing potential reaches its maximum absolute value for distances $r\sim 0.25$~fm. Moreover, it is a reasonable value for the distance in which the multipole expansion breaks down and thus for the transition from the short-distance regime to the long-distance one.

\begin{figure}[ht]
\centering
\includegraphics[width=0.6\linewidth]{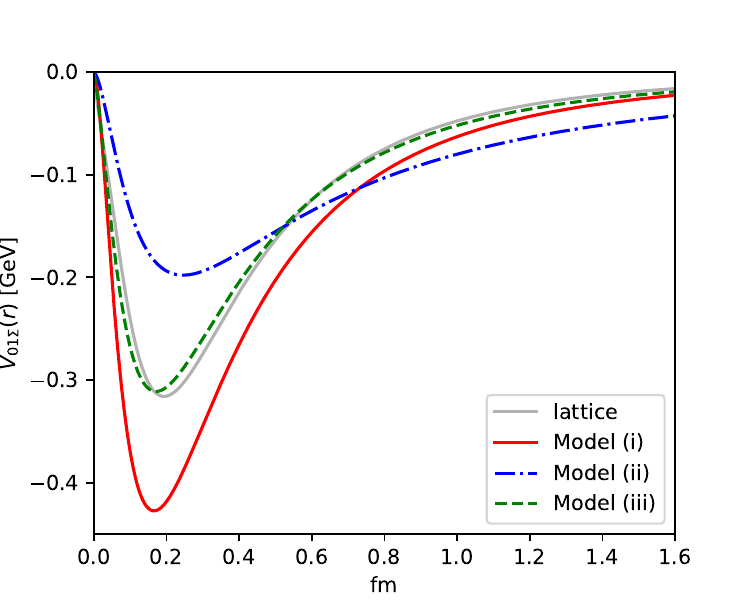} 
\caption{Plot of the three models for the quarkonium hybrid and heavy-meson pair coupling potential. The short-distance parametrization is $\sim r$ as determined from weakly-coupled pNRQCD. In the first two models the long-distance pieces behave as $r^{-3}$ and $r^{-2}$, respectively, while all the parameters values are determined by dimensional analysis. In the third model the parameters are tuned, within natural sizes, to match the quarkonium and heavy meson-antimeson pair coupling potential obtained from the lattice computation of Ref.~\cite{Bali:2005fu}, which we use as a reference for our models. The latter corresponds to the line plotted in light gray.}
\label{model_pot}
\end{figure}

We consider three different variations of the model potential. Models (i) and (ii) take the parameter values given by dimensional analysis in Eq.~\eqref{dim_an} and $n_l=3$ and $n_l=2$, respectively. Finally, model (iii) is tuned to roughly match the lattice quarkonium-heavy-meson pair mixing potential to serve as a reference point. Its parameters read as $c_s =8\Lambda^2_{\rm QCD}$, $c_l=0.9\Lambda^{-2}_{\rm QCD}$, $n_l=3$ and $r_0=0.275$~fm. In Fig.~\ref{model_pot} we plot the three model potentials and the quarkonium-heavy-meson pair mixing potential fitted in Ref.~\cite{TarrusCastella:2022rxb} to the lattice data of Ref.~\cite{Bali:2005fu}. The heavy meson-antimeson pair interpolating operator in Ref.~\cite{Bali:2005fu} is isospin $I=0$ corresponding to the normalized sum of the charged and neutral heavy meson-antimeson pairs. As we use this data as reference, we should also take ${\cal M}_{0^{-+}}$ to be isospin $I=0$ and therefore the results obtained from this model potential will correspond to the sum of neutral and charged heavy-meson pair contributions. Since the light-quarks in Ref.~\cite{Bali:2005fu} are degenerate it is also possible to take the field ${\cal M}_{0^{-+}}$ as carrying a single $q\bar{q}$ light-quark flavor and rescale the potential by $1/\sqrt{2}$. We can make use of this to obtain the coupling potential to $D_s^{(*)}\bar{D}_s^{(*)}(B_s^{(*)}\bar{B}_s^{(*)})$ in the $SU(3)$ flavor symmetry limit.

The rest of the details of the computation are as follows. The hybrid quarkonium masses and wave functions are obtained following Ref.~\cite{Berwein:2015vca} with potentials fitted to the lattice data of Ref.~\cite{Juge:2002br}. The heavy-quark masses are taken in the $RS'$ scheme at $\nu_f=1$~GeV: $m_c=1.496$~GeV and $m_b=4.885$~GeV~\cite{Peset:2018ria}. The energy offset of the lattice data is determined as the difference between the conventional $2s$ quarkonium spin-average experimental masses and the prediction obtained from the $\Sigma_g^+$ potential fitted to the lattice data of Ref.~\cite{Juge:2002br} plus the $1/m_Q$ spin-independent potential contribution computed in perturbation theory. The $1/m_Q$ spin-independent potential is fitted to  the lattice data of Refs.~\cite{Koma:2006si,Koma:2007jq}. The heavy meson masses at leading order in the heavy-quark mass expansion can be approximated by the spin averages of the physical heavy-meson states masses taken from the PDG~\cite{Workman:2022ynf}. We find
\begin{align}
m_D&=1.97322~{\rm GeV}\,,\\
m_B&=5.3134~{\rm GeV}\,,\\
m_{D_s}&=2.0762~{\rm GeV}\,,\\
m_{B_s}&=5.4033~{\rm GeV}\,.
\end{align}

\begin{table}[ht]
\centering\resizebox{\textwidth}{!}{%
\begin{tabular}{cccccccccccccc} \hline\hline
 $n\,l_\ell$ & $M_c^{(0)}$ & $\Gamma_{(i)}$ & $\Gamma_{(ii)}$ & $\Gamma_{(iii)}$ & $\delta E_{(i)}$ & $\delta E_{(ii)}$ & $\delta E_{(iii)}$ & $\Gamma^s_{(i)}$ & $\Gamma^s_{(ii)}$ & $\Gamma^s_{(iii)}$ & $\delta E^s_{(i)}$ & $\delta E^s_{(ii)}$ & $\delta E^s_{(iii)}$ \\ \hline
 $1s_0$ & $4554$ &  $7$ & $<1$ &  $5$ &  $58$ &  $31$ &  $37$ & $19$ &  $4$ & $14$ &  $45$ & $27$ & $30$ \\
 $2s_0$ & $4996$ &  $2$ & $<1$ &  $1$ &  $33$ &  $15$ &  $20$ &  $5$ & $<1$ &  $3$ &  $23$ & $11$ & $14$ \\
 $3s_0$ & $5388$ & $<1$ &  $1$ & $<1$ &  $21$ &   $8$ &  $12$ &  $1$ & $<1$ & $<1$ &  $14$ &  $5$ &  $8$ \\
 $4s_0$ & $5747$ & $<1$ &  $1$ & $<1$ &  $15$ &   $6$ &   $8$ & $<1$ & $<1$ & $<1$ &   $9$ &  $3$ &  $5$ \\
 $5s_0$ & $6081$ & $<1$ &  $2$ & $<1$ &  $11$ &   $4$ &   $6$ & $<1$ & $<1$ & $<1$ &   $6$ &  $2$ &  $4$ \\
 $1(s\backslash d)_1$ & $4110$ & $36$ & $28$ & $24$ & $-29$ & $-13$ & $-17$ &  n/a &  n/a &  n/a &  $-9$ & $-5$ & $-5$ \\
 $2(s\backslash d)_1$ & $4462$ & $49$ & $23$ & $29$ &  $-9$ &  $-1$ &  $-5$ & $20$ & $11$ & $13$ & $-10$ & $-4$ & $-6$ \\
 $3(s\backslash d)_1$ & $4765$ &  $6$ & $<1$ &  $3$ &  $20$ &  $14$ &  $13$ &  $8$ &  $2$ &  $5$ &  $14$ & $11$ &  $9$ \\
 $4(s\backslash d)_1$ & $4814$ & $28$ & $14$ & $17$ &  $-8$ &  $-1$ &  $-4$ & $11$ &  $6$ &  $7$ &  $-6$ & $-2$ & $-4$ \\
 $5(s\backslash d)_1$ & $5130$ & $23$ &  $7$ & $13$ &   $5$ &   $3$ &   $3$ & $14$ &  $5$ &  $8$ &   $2$ &  $1$ &  $1$ \\
 $1(p\backslash f)_2$ & $4321$ & $24$ & $22$ & $16$ &  $-8$ &  $-3$ &  $-5$ &  $5$ &  $7$ &  $4$ &  $-6$ & $-5$ & $-4$ \\
 $2(p\backslash f)_2$ & $4658$ & $30$ & $19$ & $19$ &  $-4$ &  $<1$ &  $-2$ & $14$ & $10$ &  $9$ &  $-5$ & $-2$ & $-3$ \\
 $3(p\backslash f)_2$ & $4965$ & $10$ &  $4$ &  $6$ &   $9$ &   $8$ &   $6$ &  $8$ &  $3$ &  $5$ &   $5$ &  $5$ &  $3$ \\
 $4(p\backslash f)_2$ & $5001$ & $16$ & $11$ & $10$ &  $-4$ &  $<1$ &  $-2$ &  $6$ &  $5$ &  $4$ &  $-3$ & $<1$ & $-2$ \\
 $5(p\backslash f)_2$ & $5297$ & $20$ &  $9$ & $12$ &   $2$ &   $2$ &   $1$ & $12$ &  $6$ &  $7$ &  $<1$ & $<1$ & $<1$ \\\hline\hline
\end{tabular}}
\caption{Contributions of $D_{(s)}^{(*)}\bar{D}_{(s)}^{(*)}$ to the widths and masses of hybrid charmonium states with $\Sigma_u^-$ component, corresponding to $H_3 [s_0]$, $H_1 [(s\backslash d)_1]$ and $H_4 [(p\backslash f)_2]$ multiplets of table~\ref{multi}.  All dimension-full entries are in MeV units. The angular momentum of the heavy meson-antimeson pair is $l'=\ell$. The subscript with Roman numbers refers to the model potential used (see Fig.~\ref{model_pot}). The superscript $s$ indicates the that the quantities refer to the heavy-meson pairs with strangeness. Energy contributions labeled as $<1$ should be understood as the absolute value of the contribution being smaller than $1$~MeV. The third column corresponds to $M^{(0)}=E^{(0)}+2m_Q$.}
\label{rslt_cc}
\end{table}

\begin{table}[ht]
\centering\resizebox{\textwidth}{!}{%
\begin{tabular}{cccccccccccccc} \hline\hline
 $n\,l_\ell$ & $M_b^{(0)}$ & $\Gamma_{(i)}$ & $\Gamma_{(ii)}$ & $\Gamma_{(iii)}$ & $\delta E_{(i)}$ & $\delta E_{(ii)}$ & $\delta E_{(iii)}$ & $\Gamma^s_{(i)}$ & $\Gamma^s_{(ii)}$ & $\Gamma^s_{(iii)}$ & $\delta E^s_{(i)}$ & $\delta E^s_{(ii)}$ & $\delta E^s_{(iii)}$ \\ \hline
 $1s_0$ & $11077$ &   $2$ &  $5$ & $<1$ & $107$ &  $51$ &  $71$ &  $3$ & $<1$ &  $6$ &  $87$ & $46$ &  $62$ \\
 $2s_0$ & $11361$ &   $1$ &  $5$ & $<1$ &  $67$ &  $28$ &  $42$ &  $4$ & $<1$ &  $4$ &  $54$ & $23$ &  $35$ \\
 $3s_0$ & $11616$ &  $<1$ &  $5$ & $<1$ &  $46$ &  $17$ &  $28$ &  $2$ & $<1$ &  $2$ &  $36$ & $14$ &  $22$ \\
 $4s_0$ & $11851$ &   $1$ &  $4$ & $<1$ &  $33$ &  $12$ &  $19$ &  $1$ & $<1$ & $<1$ &  $25$ &  $9$ &  $15$ \\
 $5s_0$ & $12071$ &   $1$ &  $4$ & $<1$ &  $25$ &   $9$ &  $15$ & $<1$ & $<1$ & $<1$ &  $19$ &  $6$ &  $11$ \\
 $1(s\backslash d)_1$ & $10783$ &  $81$ & $55$ & $64$ & $-59$ & $-14$ & $-28$ &  n/a &  n/a &  n/a & $-15$ & $-7$ & $-10$ \\
 $2(s\backslash d)_1$ & $10980$ & $129$ & $45$ & $74$ & $-21$ &  $<1$ &  $-8$ & $36$ & $18$ & $25$ & $-29$ & $-9$ & $-15$ \\
 $3(s\backslash d)_1$ & $11176$ &  $71$ & $18$ & $41$ &  $34$ &  $15$ &  $20$ & $58$ & $18$ & $35$ &  $15$ &  $9$ &   $9$ \\
 $4(s\backslash d)_1$ & $11219$ &  $14$ & $11$ &  $8$ &  $25$ &  $21$ &  $17$ & $<1$ & $<1$ & $<1$ &  $26$ & $18$ &  $18$ \\
 $5(s\backslash d)_1$ & $11380$ &  $63$ & $17$ & $35$ &   $6$ &   $3$ &   $4$ & $36$ & $11$ & $20$ &  $-3$ & $<1$ &  $-2$ \\
 $1(p\backslash f)_2$ & $10895$ &  $63$ & $39$ & $40$ & $-19$ &  $-4$ & $-10$ &  $4$ &  $6$ &  $4$ & $-14$ & $-8$ &  $-9$ \\
 $2(p\backslash f)_2$ & $11092$ &  $87$ & $40$ & $52$ & $-13$ &  $<1$ &  $-6$ & $29$ & $17$ & $19$ & $-17$ & $-7$ & $-10$ \\
 $3(p\backslash f)_2$ & $11284$ &  $75$ & $27$ & $43$ &   $9$ &   $7$ &   $6$ & $44$ & $18$ & $26$ &  $-2$ &  $1$ &  $<1$ \\
 $4(p\backslash f)_2$ & $11346$ &   $3$ &  $4$ &  $2$ &  $20$ &  $18$ &  $13$ & $<1$ & $<1$ & $<1$ &  $19$ & $14$ &  $12$ \\
 $5(p\backslash f)_2$ & $11480$ &  $64$ & $23$ & $37$ &  $<1$ &   $2$ &  $<1$ & $32$ & $13$ & $19$ &  $-6$ & $-1$ &  $-3$ \\ \hline\hline
\end{tabular}}
\caption{Contributions of $B_{(s)}^{(*)}\bar{B}_{(s)}^{(*)}$ to the widths and masses of hybrid bottomonium states with $\Sigma_u^-$ component, corresponding to $H_3 [s_0]$, $H_1 [(s\backslash d)_1]$ and $H_4 [(p\backslash f)_2]$ multiplets of table~\ref{multi}. All dimension-full entries are in MeV units. The angular momentum of the heavy meson-antimeson pair is $l'=\ell$. The subscript with Roman numbers refers to the model potential used (see Fig.~\ref{model_pot}). The superscript $s$ indicates that the quantities refer to the heavy-meson pairs with strangeness. Energy contributions labeled as $<1$ should be understood as the absolute value of the contribution being smaller than $1$~MeV. The third column corresponds to $M^{(0)}=E^{(0)}+2m_Q$.}
\label{rslt_bb}
\end{table}

We display our results for the energy shifts caused by the heavy-meson pairs to the hybrid masses and the decay widths of the hybrid states into the heavy-meson pairs in tables~\ref{rslt_cc} and \ref{rslt_bb} for charmonium and bottomonium hybrids, respectively. In the charmonium sector we find that the energy contributions are, with a few exceptions, below the $30$~MeV, thus being smaller than the $\mathcal{O}(1/m_Q)$ spin-dependent operator contributions~\cite{Brambilla:2018pyn,Brambilla:2019jfi,Soto:2023lbh} and also smaller than the expected size of the spin-independent ones, which so far have not been computed. The energy contributions tend to decrease with increasing $n$ and $\ell$. The widths range between $1$ to $20$~MeV per channel\footnote{Notice, that neutral and charged $D$ meson contributions are added up in columns 4 to 9 of table \ref{rslt_cc}.}. The larger widths are found in states closer to the heavy-meson pair thresholds, with the values being very sensitive to the energy gap between the state and threshold. Adding up the widths for the two channels studied, we find values that range from slightly larger than the semi-inclusive widths for transitions to quarkonium~\cite{Oncala:2017hop,Brambilla:2022hhi} to much smaller ones. The latter, however, correspond to higher lying hybrid charmonium states that can also decay to higher mass heavy-meson pairs not considered here.

In the bottomonium sector, the energy contributions are larger than in the charmonium sector, with up to ten states having contributions larger than $30$~MeV. As a consequence for bottomonium hybrids the $B_{(s)}^{(*)}\bar{B}_{(s)}^{(*)}$ contribution to the mass is significantly larger than the spin-dependent one. The widths are a factor of $2$ to $3$ larger than in the charmonium sector. The larger contributions in the bottomonium sector can be traced back to the fact that the coupling operator is not heavy-quark mass suppressed and the less extended hybrid bottomonium wave functions having a larger overlap with the coupling potential than the more extended wave functions of hybrid charmonia. In contrast, semi-inclusive transition widths to quarkonium are smaller for bottomonium hybrids than for charmonium hybrids~\cite{Oncala:2017hop,TarrusCastella:2021pld,Brambilla:2022hhi} and consequently also smaller than decay widths to $B_{(s)}^{(*)}\bar{B}_{(s)}^{(*)}$. As in the charmonium sector, the lowest mass $s_0$ states have the largest contributions to the mass, while the larger widths are found in the lowest lying $(s\backslash d)_1$ states. 

It is perhaps unexpected that the widths for $s_0$ are smaller than those of other states of similar mass, while the energy shifts are larger. This can be explained as follows. The transition form factors into $S$-wave heavy-meson pairs, $a_{n0;S}(k)$, have the largest peak, resulting in large values for the energy shifts produced by the heavy-meson pairs, however, they decrease very rapidly to zero as $k$ increases. As the $s_0$ states are far away from the thresholds the values of $p_r$ are large enough for the widths to correspond to the region where the form factors are small. Moreover, as the energy contribution is overall positive, it tends to increase the value of hybrid mass and, correspondingly, of $p_r$, further decreasing the value of the width. 

In table~\ref{h2h5} we show the spectra of the pure $\Pi_u$ states that do not couple with the two lowest lying thresholds at leading order. The lowest mass states ($1p_1$), in both sectors, are interesting since the decays into $D_{(s)}^{(*)}\bar{D}_{(s)}^{(*)}(B_{(s)}^{(*)}\bar{B}_{(s)}^{(*)})$ are $1/m_Q$-suppressed and are below any other heavy-meson pair threshold. Furthermore, two of the spin-triplet states of corresponding multiplet have explicitly exotic $j^{pc}$. If they could be measured, identifying these states as hybrids should be straightforward due to these two very distinct characteristics.

\begin{table}[ht]
\centering
\begin{tabular}{ccc} \hline\hline
 $n\,l_\ell$ &  $M^{(0)}_{c}$ &  $M^{(0)}_{b}$  \\ \hline
 $1p_1$ & $4238$ & $10840$ \\
 $2p_1$ & $4610$ & $11059$ \\
 $3p_1$ & $4954$ & $11267$ \\
 $4p_1$ & $5277$ & $11466$ \\
 $5p_1$ & $5583$ & $11657$ \\
 $1d_2$ & $4429$ & $10945$ \\
 $2d_2$ & $4786$ & $11160$ \\
 $3d_2$ & $5118$ & $11364$ \\
 $4d_2$ & $5433$ & $11560$ \\
 $5d_2$ & $5732$ & $11748$ \\ \hline\hline
\end{tabular}
\caption{Masses of the pure $\Pi_u$ hybrid quarkonium states that do not couple to $D_{(s)}^{(*)}\bar{D}_{(s)}^{(*)}(B_{(s)}^{(*)}\bar{B}_{(s)}^{(*)})$ corresponding to the $H_2 [p_1]$ and $H_5 [d_2]$ multiplets of table~\ref{multi}. The masses are obtained as $M^{(0)}=E^{(0)}+2m_Q$ and are given in MeV units.}
\label{h2h5}
\end{table}

\begin{figure}[ht]
\centering
\includegraphics[width=0.5\linewidth]{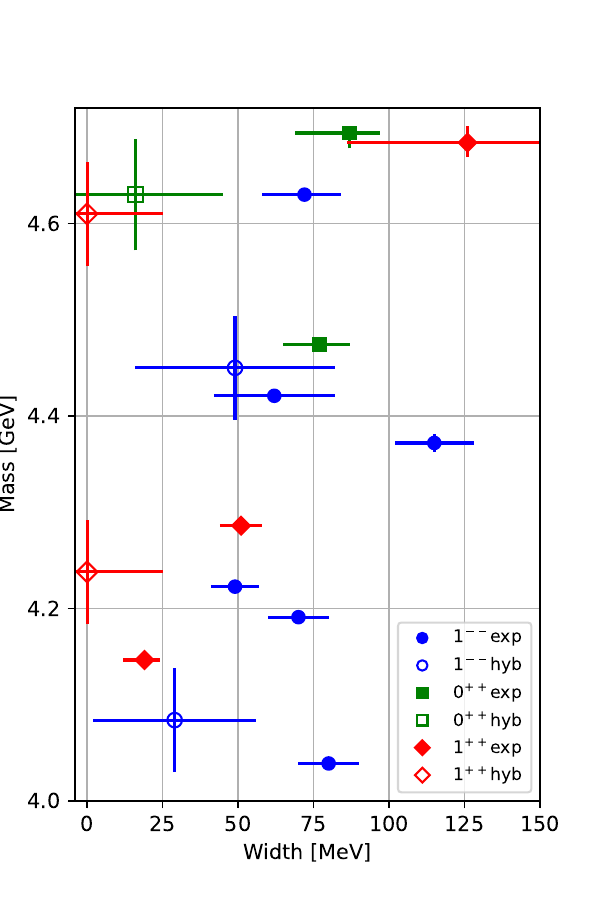}
\caption{Comparison of the mass and widths of hybrid charmonium states in tables~\ref{rslt_cc} and \ref{h2h5} with experimental states above threshold listed in the PDG~\cite{Workman:2022ynf}. The masses and widths of the hybrid states are obtained as the average of the results corresponding to the three model potentials in Fig.~\ref{model_pot}. The bars correspond to the sum in quadrature of the standard deviation of the results for the three models and the size of higher order corrections $\mathcal{O}(\Lambda_{\rm QCD}/m_c)$.}
\label{comp_exp_cc}
\end{figure}

\begin{figure}[ht]
\centering
\includegraphics[width=0.5\linewidth]{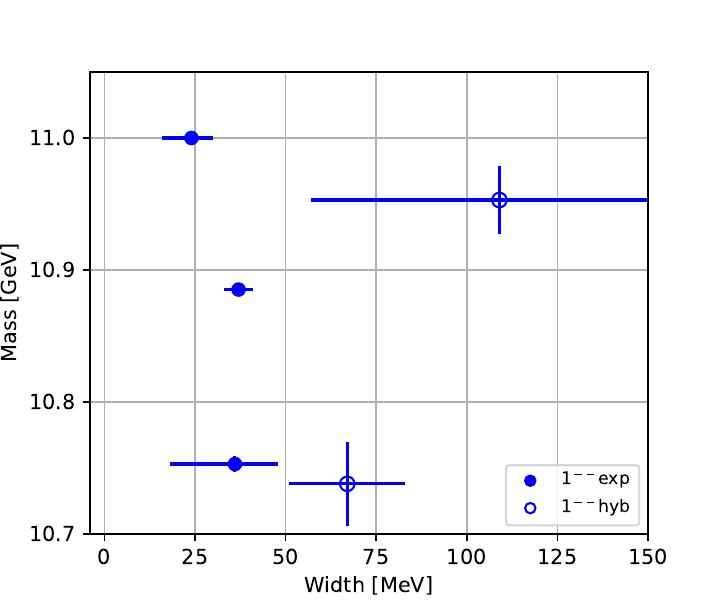}
\caption{Comparison of the mass and widths of hybrid bottomonium states in table~\ref{rslt_bb} with experimental states above threshold listed in the PDG~\cite{Workman:2022ynf}. The masses and widths of the hybrid states are obtained as the average of the results corresponding to the three model potentials in Fig.~\ref{model_pot}. The bars correspond to the sum in quadrature of the standard deviation of the results for the three models and the size of higher order corrections $\mathcal{O}(\Lambda_{\rm QCD}/m_b)$.}                                                      
\label{comp_exp_bb}
\end{figure}

In Figs.~\ref{comp_exp_cc} and \ref{comp_exp_bb} we compare the average of our results for the three model potentials, respectively, with the experimental charmonium and bottomonium spectra above threshold. In the charmonium sector states with three distinct $j^{pc}$ have been observed. These are, six $1^{--}$ states, three $1^{++}$ and two $0^{++}$. From table~\ref{multi} one can find that the hybrid states with matching quantum numbers are: the spin singlet of the $(s\backslash d)_1$ multiplet for $1^{--}$; the spin singlet of the $p_1$ multiplet for $1^{++}$; and the spin singlet of the $s_0$ multiplet for $0^{++}$. The candidates with compatible masses are $\psi(4160)$ and $\psi(4415)$, for $n=1$ and $n=2$ of $(s\backslash d)_1$, and  $\chi_{c1}(4274)$ and $\chi_{c0}(4700)$ for the ground state of $p_1$ and the $n=2$ of $s_0$, respectively. The experimental widths of these are smaller than our model average, which leaves room for additional contributions, such as the semi-inclusive transition widths computed in Refs.~\cite{Oncala:2017hop,Brambilla:2022hhi}. Adding the widths of Ref.~\cite{Brambilla:2022hhi} to the ones we computed in the present work results in larger widths than the experimental ones although compatible considering the uncertainties.

In the bottomonium sector there are only three observed states above threshold, shown in Fig~\ref{comp_exp_bb}, all with quantum numbers $1^{--}$. There are two hybrid bottomonium states in the mass region with matching quantum numbers, the $n=1,2$ spin-singlets of the $(s\backslash d)_1$ multiplets. In particular, the $n=1$ state has a predicted mass compatible with $\Upsilon(10753)$. However, the widths we have obtained are consistently larger than the ones of the experimental states, moreover this is before we consider additional decay channels. Therefore, with the present results it is unlikely that the bottomonium states above threshold so far observed are quarkonium hybrids.

\section{Conclusions}\label{sec:con}

We have presented the leading order BOEFT couplings of double-heavy hadrons to pairs of heavy hadrons for any spin of the light degrees of freedom. We have shown how to transform the expression of the  BOEFT states from using a basis of fields in a standard spherical spin representation to a basis of fields belonging to $D_{\infty h}$ representations. In this base the coupling becomes extremely simple and the computation of its expected value between a double-heavy hadron and a heavy hadron pair states is straightforward. Using the expected value of the coupling, the contribution of the heavy-hadron pairs to the self-energy of the double-heavy hadrons can be computed, and from it the effect of the former on the mass and width of the latter. The matching of the potential accompanying the coupling operator to a NRQCD correlator has not been presented. However, it can be straightforwardly obtained following the matching computation for the coupling potential for conventional quarkonium to a heavy-meson pair presented in Ref.~\cite{TarrusCastella:2022rxb}, by replacing the quarkonium interpolating operator for the double-heavy hadron one, which can be found in Ref.~\cite{Soto:2020xpm}.

In Section~\ref{s:hqdm} we have applied the results of Section~\ref{s:ths} for the case of the coupling of $D_{(s)}^{(*)}\bar{D}_{(s)}^{(*)}(B_{(s)}^{(*)}\bar{B}_{(s)}^{(*)})$ to hybrid quarkonium. In particular, we have studied the hybrid states generated by the two lowest static energies, $\Sigma_u^-$ and $\Pi_u$. From the computation of the expected value of the coupling operator we have found selection rules for these decays. The first one is that a hybrid state without a $\Sigma_u^-$ component cannot decay through the leading order coupling operator. Secondly, only a single angular momentum, equal to the hybrid total angular momentum quantum number $\ell$, is allowed for the heavy-meson pair product of the decay. The $1p_1$ states, both the in charm and bottom sectors, turn out to be very interesting. Due to being pure $\Pi_u$, these states have $1/m_Q$-suppressed decays to $D_{(s)}^{(*)}\bar{D}_{(s)}^{(*)}(B_{(s)}^{(*)}\bar{B}_{(s)}^{(*)})$ and, due to their leading order mass, they are not kinematically allowed to decay to any other heavy-meson pair. Furthermore, two of the $j^{pc}$ states in the corresponding heavy quark spin-symmetry multiplets are explicitly exotic, thus being good candidates to be unambiguously identified as hybrids.

The potential in the coupling operator must be computed with nonperturbative techniques and so far is unknown. We have built a model for it based on its expected short- and long-distance behaviors and inspired by the analogous potential coupling conventional quarkonium to heavy-meson pairs for which there exists lattice data. For three variations of this model potential we have obtained the energy shifts caused by the heavy-meson pairs to the hybrid masses as well as the decay widths of the hybrid states into the heavy-meson pairs. We presented our results in tables~\ref{rslt_cc} and \ref{rslt_bb}.
A remarkable result is that the energy shifts and widths are larger in the bottomonium sector than in the charmonium one. This is a result of the coupling operator in Eq.~\eqref{mix_lgrngn} not being heavy-quark mass suppressed and the potentials (in Fig.~\ref{model_pot}) having the larger values in the short-distances. A comparison with the experimental quarkonium states above open flavor threshold can be found in Figs.~\ref{comp_exp_cc} and \ref{comp_exp_bb}. In the charmonium sector $\psi(4160)$, $\psi(4415)$, $\chi_{c1}(4274)$ and $\chi_{c0}(4700)$ remain good candidates for hybrid states. In the bottomonium sector the hybrid states turn out to have too large widths to be compatible with the experimental states observed so far.

%\section*{Acknowledgments}                             

\appendix

\bibliographystyle{JHEP}
\bibliography{hhmbib}

\end{document}